\documentstyle[epsfig,twocolumn]{esapub}

\begin{document}

\setlength{\parindent}{0pt}
\setlength{\parskip}{ 10pt plus 1pt minus 1pt}
\setlength{\hoffset}{-1.5truecm}
\setlength{\textwidth}{ 17.1truecm }
\setlength{\columnsep}{1truecm }
\setlength{\columnseprule}{0pt}
\setlength{\headheight}{12pt}
\setlength{\headsep}{20pt}

\pagestyle{esapubheadings}
\title{CONSISTENT SOLAR MODELS INCLUDING THE $^7$Li AND $^3$He
CONSTRAINTS}

\author{{\bf Sylvie Vauclair and Olivier Richard} \\
Laboratoire d'Astrophysique, Observatoire Midi-Pyr\'en\'ees \\
14, Av. E. Belin, 31400 TOULOUSE, FRANCE}

\maketitle

\begin{abstract}

Improvements induced by helioseismology show that microscopic diffusion
cannot be neglected in solar models. Helium and metal diffusion is now
generally introduced in the computations. They cannot however explain
the observed lithium depletion. Macroscopic motions are needed to bring
up to the convection zone the nuclearly depleted matter. Such motions
also lead to a better agreement between the sound velocity in the model
and that deduced from the inversion of seismic modes. On the other hand,
they must not bring up too much $^3$He, as we now know that the $^3$He/$^4$He
ratio cannot have changed by more than 10 percent during these last 3
Gyr.
Here we present new consistent solar models in which the $^3$He
constraint is satisfied. The fact that $^3$He must remain constant while
$^7$Li must be depleted leads to strong constraints on the effective
macroscopic diffusion coefficient allowed below the solar convection
zone.
\vspace {5pt} \\

  Key~words: Sun : abundances; helioseismology, diffusion processes

\end{abstract}

\section{SETTLING AND MIXING}

The importance of helium and heavy element settling inside the Sun is
now widely recognized. Solar models computed in the old ``standard'' way,
in which the element settling is totally neglected, do not agree with
the inversion of the seismic modes. The discrepancy is much reduced when
element settling is introduced. This result has been obtained by many
authors, in different ways (see Gough et al. (1996) and references
therein).

For the first time, Turcotte et al. (1998) have consistently computed the
radiative accelerations on the elements included in the OPAL opacities.
They have found that, contrary to current belief, the effect of
radiation can, in some cases, be as large as $\cong~40\%$ that of
gravity below the solar convective zone. This is important only for
metals however, and not for helium. When the radiative accelerations are
neglected, the abundances of most metals change by about $\cong~7.5\%$
if complete ionization is assumed below the convective zone, and by
$\cong~8.5\%$ if detailed ionization rates are computed. When the
radiative accelerations are introduced, with detailed ionization, the
results lie in-between.

Although the introduction of pure element settling in the solar models
considerably improve the consistency with the seismic Sun, some
discrepancies do remain, particularly below the convective zone where a
``spike'' appears in the sound velocity (Guzik 1998, Turck-Chi\`eze et al.
1998). It has been shown that this behavior may be due to the helium
gradient which would be too strong in case of pure settling. Mild
macroscopic motions below the convection zone slightly decrease this
gradient and helps reducing the discrepancy (Richard et al. 1996 - RVCD,
Corbard et al. 1998). The helium profiles directly obtained from
helioseismology (Basu 1997, Antia and Chitre 1997) show indeed a helium
gradient smoother than the gradient obtained with pure settling.

\section{LIGHT ELEMENT ABUNDANCES}

The abundance determinations in the solar photosphere show that lithium
has been depleted by a factor of about 140 compared to the protosolar
value while beryllium
is generally believed to be depleted by a factor 2.
These values have widely
been used to constraint the solar models (e.g. RVCD).
However, while the lithium depletion factor seems well established, the
beryllium value is still being discussed. Balachandran and Bell
(1998) argue that the beryllium depletion is not real because of
insufficient inclusion of continuous opacity in the abundance
determination. Their new treatment leads to a solar value identical to
the
meteoritic value.

Observations of the $^3$He/$^4$He ratio in the solar
wind
and in the lunar rocks (Geiss 1993,
Gloecker and Geiss 1996, Geiss and Gloecker 1998) show that this ratio
may not
have increased by more than $\cong~10\%$ since 3 Gyr in the Sun, which
is in contradiction with the results of RVCD.

While the occurrence of some mild mixing below
the solar convection zone is needed to explain
the lithium depletion and helps for the conciliation of the models with
helioseismological constraints,
the $^3$He/$^4$He
observations put a strict constraint on its efficiency. The solar
chemical composition appears as a powerful tool in order to produce
consistent solar models.

\section{CONSTRAINTS ON THE MIXING BELOW THE CONVECTION ZONE}

In RVCD, a mild mixing below the convection zone, attributed to
rotation-induced shears (Zahn 1992), was introduced. This mixing was
supposed to be stopped at some depth due to a stabilizing mu-gradient.
Such an assumption is necessary to prevent mixing in the nuclear core
regions, otherwise we do not obtain the right sound velocity
(Richard and Vauclair, 1997; Bahcall et al., 1997; 
Degl'Innocenti and Ricci, 1998). 
The best model presented in RVCD was obtained assuming
a critical mu-gradient of $4. 10^{-13}$ cm$^{-1}$,
compatible with the order of magnitude deduced from Huppert and Spiegel
1977.
In this model, lithium was depleted
by a factor 140 and beryllium by about 2.  
On the other hand the $^3$He abundance increased too much in the
convection zone compared to the recent determinations.

We now know that there may be a lithium depletion by a factor 3 
in the pre-main sequence evolution, 
and that beryllium may not be depleted at all, which relaxes the
constraints. The most stringent constraint is now given by 
the $^3$He determinations.

We have tried several parametrizations of mixing below the solar
convection zone, which could reproduce both the $^7$Li and the $^3$He
constraints. The only way to obtain such a result is to postulate a mild
mixing, which would be efficient down to the lithium nuclear burning
region but not too far below, to preserve the original $^3$He abundance.
A mixing effect decreasing with time, as obtained with the
rotation-induced shear hypothesis, helps satisfying the constraints, as
the $^3$He peak itself builts up during the solar life.

Here we present results obtained with a similar prescription as in RVCD
(rotation-induced mixing) in which the critical mu-gradient 
($\nabla \mu_c$) 
was allowed
to vary between $1. 10^{-13}$ and $5. 10^{-13}$cm$^{-1}$.

\begin{figure}[h]
  \begin{center}
    \epsfxsize=.95\hsize
    \epsffile{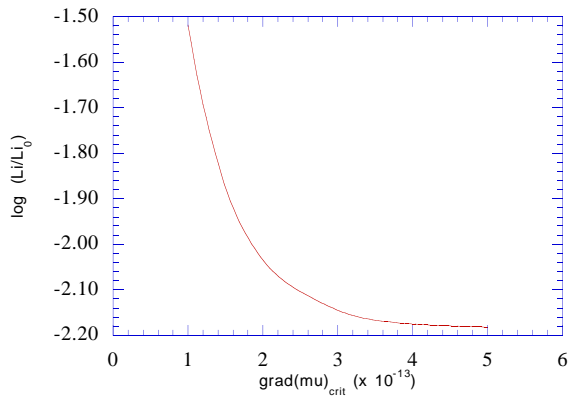}
        \caption{Lithium depletion at the solar age as a function of
    $\nabla \mu_c$}
    \label{fig:li}
  \end{center}
\end{figure}

Figure 1 shows the main-sequence lithium depletion obtained at the solar
age as a function of $\nabla \mu_c$. If we take into account
the depletion during the pre-main sequence, lithium has to be
depleted by at least 30 during the main-sequence to account for the
observations. From Figure 1, $\nabla \mu_c$ must be larger than
$1. 10^{-13}$ cm$^{-1}$.

Figure 2 displays the $^3$He/$^4$He increase with time for several
values of $\nabla \mu_c$. The observational constraints lead
to $\nabla \mu_c \leq 2. 10^{-13}$ cm$^{-1}$. We thus find that the
observations of the light elements in the Sun may be accounted for with
a rotation-induced mixing limited by a mu-gradient of 
$(1.5 \pm 0.5). 10^{-13}$ cm$^{-1}$.

\begin{figure}[h]
  \begin{center}
    \epsfxsize=.95\hsize
    \epsffile{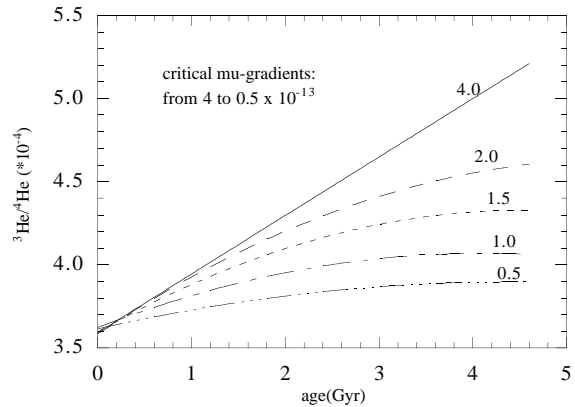}
        \caption{$^3$He/$^4$He variation with time for various $\nabla
    \mu_c$ (in units of $10^{-13}$ cm$^{-1}$)}
    \label{fig:he3}
  \end{center}
\end{figure}

The corresponding shape of the effective diffusion coefficient is given
in Figure 3. It is interesting to point out the similarity between this
result and the order of magnitude of the diffusion coefficient 
needed to smooth out
the spike in the sound velocity compared to the inversion of heliosismic
modes (Morel et al., 1998).

\begin{figure}[ht]
  \begin{center}
    \epsfxsize=.95\hsize
    \epsffile{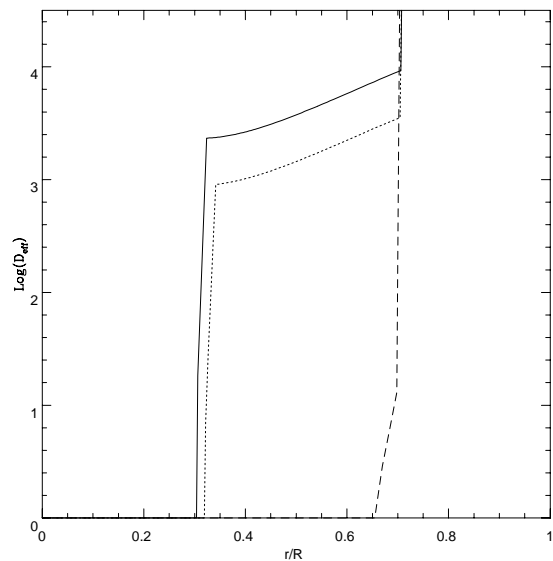}
    \caption{Effective diffusion coefficient at various ages for a
    model with $ \nabla \mu_c = 1. 10^{-13}$. The
    coefficient decreases rapidly with age (full line for .22 Gyr,
    short dashed line for .42 Gyr, and dashed line for 4.6 Gyr).}
    \label{fig:dt}
  \end{center}
\end{figure}

\section{COMPARISON WITH OTHER STARS}

The lithium abundance in the Sun may be compared to that observed in
galactic clusters of known ages 
($\alpha $Per: Balachandran et al., 1996; Pleiades: Soderblom et
al., 1993a; UMaG: Soderblom et al., 1993b;
 Hyades: Thorburn et al., 1993; NGC752: Balachandran,
1995).
Following the solar evolutionary
track we may attribute an effective temperature and, from the
observations, a lithium abundance for each age.
Figure 4 displays the observed lithium variations with age and the
theoretical ones obtained with 
$\nabla \mu_c$ = $1.$ and $2. 10^{-13}$cm$^{-1}$.
We can see that the parametrization of the mixing below the
convection zone, which best accounts for the solar
light element abundances and helioseismology, also
correctly reproduce the lithium abundance evolution with time.

\begin{figure}[h]
  \begin{center}
    \epsfxsize=.95\hsize
    \epsffile{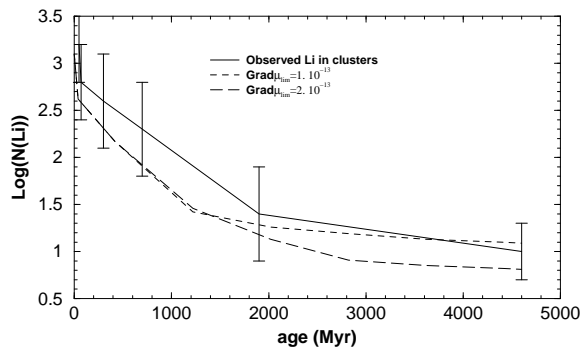}
        \caption{Lithium depletion with time in 1$M_{\odot}$ stars obtained
    from the observations of galactic clusters and comparison with the two
    theoretical models}
    \label{fig:gradmu}
  \end{center}
\end{figure}

\section{SUMMARY}

In summary the best solar models must include the effect of element
settling, which represents an improvement on the physics, without any
free parameter added. These models can be considered as the new
``standard'' models. They cannot however reproduce the $^7$Li depletion
and they lead to a spike in the sound velocity, compared to the seismic
Sun, just below the convective zone. These two observations suggest the
presence of some mild mixing in this region of the internal Sun. Adding
the constraint on the abundance of $^3$He as given by Geiss
and Gloecker (1998) leads to a precise description of the allowed
profile of the macroscopic diffusion coefficient below the convective
zone. The result is consistent with the mixing needed to smooth out the
spike in the sound velocity and it leads to a lithium abundance
variation with time consistent with the observations of galactic
clusters.
This parametrized description can be taken as a 
challenge for the hydrodynamicists.

{}


\begin{thebibliography}{}

\bibitem{}Antia, H.M., Chitre, S.M.. 1997, {\it astro-ph/9710159}



\bibitem{}Balachandran, S. 1995, {\it ApJ} {\bf 446}, 203

\bibitem{}Balachandran, S., Lambert, D.L., Stauffer, J.R. 1996, {\it
    ApJ} {\bf 470}, 1243

\bibitem{}Balachandran, S., Bell, R.A.  1998, preprint

\bibitem{}Bahcall, J.N., Pinsonneault, M.H., Basu, S.,
  Christensen-Dalsgaard, J. 1997, {\it Phys. Rev. Lett.} {\bf 78}, 171 

\bibitem{}Basu, S. 1997, {\it Mon. Not. R. Astron. Soc.} {\bf 288},
572






\bibitem{}Corbard, T., Berthomieu, G., Provost, J., Blanc-F\'eraud,L.
1998, These Proceedings


\bibitem{}Degl'Innocenti, S., Ricci, B. 1998, {\it Astroparticle
    Physics} {\bf 8}, 293



\bibitem{}Geiss, J. 1993, {\it Origin and Evolution of the Elements},
ed. Prantzos, Vangioni-Flam \& Cass\'e (Cambridge Univ.
Press), {\bf 90}

\bibitem{}Gloecker, G., Geiss, J. 1996, {\it Nature} {\bf 381},
  210

\bibitem{}Geiss, J., Gloecker, G. 1998, {\it Space Sci. Rev.}, in press

\bibitem{}Gough,D.O., Kosovichev, A.G., Toomre, J., Anderson, E., Antia,
H.M., Basu, S.,  Chaboyer, B., Chitre, S.M., Christensen-Dalsgaard, J.,
Dziembowski, W.A., Eff-Darwich, A., Elliott, J.R., Giles, P.M., Goode,
P.R., Guzik, J.A., Harvey, J.W., Hill, F., Leibacher, J.W., Monteiro,
M.J.P.F.G., Richard, O., Sekii, T., Shibahashi; H., Takata, M.,
Thompson, M.J., Vauclair, S., Vorontosov, S.V. 1996, {\it Science}
{\bf 272}, 1296


\bibitem{}Guzik, J.A. 1998, These Proceedings

\bibitem{}Hupper, H.E., Spiegel, E.A. 1977, {\it ApJ} {\bf 213}, 157




\bibitem{}Morel, P., Provost, J., Berthomieu, G. 1998, These Proceedings







\bibitem{}Richard, O., Vauclair, S., Charbonnel, C., Dziembowski,
W.A. 1996, {\it A\&A} {\bf 312}, 1000

\bibitem{}Richard, O., Vauclair, S. 1997, {\it A\&A} {\bf 322}, 671


\bibitem{} Soderblom, D.R., Jones, B.F., Balachandran, S., Stauffer,
  J.R., Duncan, D.K., Fedele, S.B., Hudon, J.D. 1993a, {\it AJ} {\bf 106},
  1059

\bibitem{} Soderblom, D.R., Pilachowski, C.A., Fedele, S.B., Jones,
  B.F. 1993b, {\it AJ} {\bf 105}, 2299


\bibitem{}Thorburn, J.A., Hobbs, L.H., Deliyannis, C.P., Pinsonneault,
  M.H. 1993, {\it ApJ} {\bf 415}, 150


\bibitem{}Turck-Chi\`eze, S., Basu,S., Berthomieu,G., Bonanno,A., Brun,
A.S., Christensen-Dalsgaard, J., Gabriel, M. , Morel, P., Provost, J.,
Turcotte, S., and the GOLF team 1998, These Proceedings

\bibitem{}Turcotte, S., Richer, J., Michaud, G., Iglesias, C.A.,
  Rogers, F.J.
1998, {\it ApJ}, in press

\bibitem{}Vauclair, S. 1998, {\it Space Sci. Rev.}, in press

\bibitem{}Vauclair, S., Vauclair, G. 1982, {\it ARA\&A} {\bf 20}, 37



\bibitem{}Zahn, J.P. 1992, {\it A\&A} {\bf 265}, 115


\end{thebibliography}
\end{document}